# On the Polarization of non-Guassian optical quantum field: higher-order optical-polarization

Ravi S. Singh[1#] and Hari Prakash[2]

[1] Department of Physics, D. D. U. Gorakhpur University, Gorakhpur- 273009, (U.P.), INDIA.
[2] Physics Department, University of Allahabad, Allahabad- 211002, (U.P.), INDIA

## Abstract

Polarization of light signifies transversal, anisotropic and asymmetrical statistical property of electromagnetic radiation about direction of propagation. Traditionally, optical-polarization is characterized by Stokes theory susceptible to be insufficient in assessing polarization structure of optical quantum fields and, also, does not decipher twin characteristic polarization parameters ('ratio of real amplitudes and difference in phases'). An alternative way, in spirit of classical description of optical-polarization, is introduced which can be generalized to deal higher-order polarization of quantum light, particularly, prepared in non-Guassian Schrodinger Cat or Cat-like states and entangled bi-modal coherent states. On account of pseudo mono-modal or multi-modal nature of such optical quantum field, higher-order polarization is seen to be highly sensitive to the basis of description.

*Introduction.* -Polarization is an intrinsic trait of paraxial optical field ensuring its vectorial transversal nature. It offers a resilient degree of freedom paving ways for showcasing novel experiments in Quantum Optics [1]. The study of polarization of a beam of light witnessed two extreme states, perfect-polarized light and unpolarized light between which infinitely many states of neither polarization exist. Traditionally, Stokes parameters, discovered by G. G. Stokes in 1852 [2] and interpreted geometrically by H. Poincare in 1892 [3] on a three dimensional sphere, quantify the states of polarization of optical field and is revered as a didactic theory in classical optics. In transiting to quantum regime these parameters acquired the roles of operators, namely, Stokes operators [4] of which quantum average values specify the states of polarization of optical quantum fields.

In 1971 rigorous investigation of quantum nature of unpolarized light is carried out [5, 6] by determining structure of density operator in Fock-state representation which characterizes its general statistical properties. Renewed interests [7] in unpolarized light garnered new insights regarding its nature and state-tomography [8]. Soon after perfect polarized state of a plane monochromatic light beam is characterized by determining its density operator in diagonal coherent-state representation [9] through demand of vanishing amplitude (signal) in, at least, one orthogonal mode after passing through a compensator followed by a rotator, i.e., through SU(2) transformations. Earliest [5] as well as recent noteworthy studies [10] on optical-polarization brought forth inadequacy of Stokes theory. Moreover, a unified second-order theory of coherence and polarization of stochastic optical field [11], accounting for the changes in states of polarization of light on propagation [12], is developed by recognizing the fact about Stokes parameters that they imbibe correlations at an arbitrary spatio-temporal point.

Scrutiny of multi-photon quantum interference experiments [13] exposed that higher-order intensity-correlations in transverse plane may exhibit anisotropic non-vanishing values revealing asymmetry about direction of propagation and, thereby, displaying polarization structure where, again, Stokes parameters mislead by assigning vanishing values to themselves. Polarization of such optical fields [14] is characterized by nine parameters arranged in Hermitian matrix which, having entries defined by correlations of the intensities in orthogonal polarization modes, underscored fourth - order polarization effects. Furthermore, it is articulated [15] that the state of polarization of bi-modal optical quantum fields can, solely, be characterized by correlation functions involving equal numbers, say N, of bosonic annihilation operators to those of creation operators of which measurements are prescribed through reconstruction of the polarization sector of the density matrix. Recently, an operational scheme [16] is proposed to measure such correlation functions by establishing that these arbitrary $N^{th}$-order correlation functions can be measured by $N^{th}$-order intensity moments. Also a judicious operational criterion [17] is laid down in terms of minimal fluctuations of Stokes parameters on the Poincare sphere which, tacitly, encompasses fourth-order correlations among optical field amplitudes. Rational observations to such higher-order correlation functions point out a fundamental issue stemming from its very definition pertaining to involvement of only equal numbers of bosonic annihilation operators to those of creation operators, a critical condition dictated by phase-insensitive optical measurements based on photon-detection probabilities [18]. Also quantum Stokes theory or its higher-order version is so far applied only to Guassian optical fields such as coherent states [19] and squeezed-vacuum states [20], where general correlation functions having unequal powers of bosonic annihilation and creation operators vanishes [21]. Although the measurement of these most general correlation functions is proposed by employing balanced homodyne correlation measurements and, also, their relevance is extensively researched to display non-classical properties of optical fields [22], their importance in studies of optical-polarization of non-Guassian optical fields received little attention. We urged that their roles get exemplified only when one attempts to seek the inherent spatio-temporal nature of characteristic polarization parameters. Furthermore, it may be noted that study of optical-polarization of non-Guassian quantum states such as non-classical Schrodinger Cat states [23] or Cat-like states [24], entangled bi-modal coherent states [25] are not yet explored and remains unaddressed.

Quite recently, by generalizing circular polarization basis of spin angular momenta to that of total optical angular momentum, analogous Stokes-parameters for vector vortex beam [26] are constructed classically to describe the state of polarization for such non-paraxial optical fields. Explicitly, such generalization presumes that Stokes parameters (operators in quantum theory) are equivalent to Jordon-Schwinger spin angular momenta [27] leading to vectorial addition to orbital angular momenta of the three-dimensional optical field. But, it is critically observed that Stokes operators find distinct sort of Hilbert space [28] for its operation vis-a-vis to that of spin-angular momenta. Such critic offers, therefore, a moot point whether Stokes parameters can be generalized in quantum regime to describe states of polarization of optical non-paraxial fields of which description witnessed some alternative proposals in the literature [29]

Numerous computable-measures, based on the abstract notion of 'distance' between quantum state in question to that of unpolarized optical field in Hilbert space [30] as well as in quantum phase space [31] are introduced giving optional expressions for degree of polarization for the same quantum light. Evidently, these approaches are well-strategized, yet it does not figure out endowed relationship between amplitudes and phases of transverse orthogonal optical field modes and, also, vitally, doesn't correspond to classical-description of optical-polarization.

Moreover, recent trends in Quantum Optics ushered new physical effects such as quantum imaging, ghost imaging and spatio-temporal multipartite entanglement [32] which can only be construed bearing in mind the spatio-temporal features of optical field, its quantum state engineering and basis-based measurements. Variances of Stokes parameters or 'special' higher-order correlations took little heed to the innate messages preserved in statistical spatio-temporal properties of the optical field. This negligence has not encountered any severe inconclusiveness regarding the of state of polarization by Stokes theory in Classical Optics and, to some extent, in Quantum Optics, when Guassian light's polarization is under consideration. But, one may posit that an alternative characterization of optical-polarization relying on imperceptible spatio-temporal variations of real amplitudes and phases is imperative for non-Guassian light of which representative example is entangled bi-modal coherent states where Stokes theory seems susceptible to ascertain its states of polarization in entire regimes [33].

Our perspective on the theory of polarization of light is rooted in firm adherence to classical-description, i.e., by the superposition of two synchronous orthogonal harmonic oscillators resembling dynamically two transverse orthogonal components of a harmonic electromagnetic plane wave preserving non-random values of 'ratio of real amplitudes' and 'difference in phases'. In quantum regime an optical-polarization operator has been worked out by invoking the fact that if light were perfectly polarized in a specific mode, signal (photons) would be absent in orthogonal mode [34]. This optical-polarization operator deciphers the 'characteristic polarization-parameters'[35] for perfect polarized light, i.e., the non-random values of the 'ratio of real amplitudes' and 'difference in phases' specifying a point on Poincare sphere without relying on quantized nature of phase [36].

We, **in this letter**, generalize theory of optical-polarization which introduces a notion of higher ($n^{th}$) order optical-polarization (HOOP) by demanding that, for a single mode paraxial optical field, the ratio of complex amplitudes in transverse orthogonal modes is random but all its multiple-exponents of some least positive integer, say, 'n' are non-random parameters. This definition of HOOP is applied to extract the characteristic polarization parameters of optical field in non-classical Schrodinger Cat or Cat-like states and entangled bi-modal coherent states. Higher-order polarized light is seen not to be a mono-modal optical field as it reveals uncanny nature that 'order of polarization' is sensitive to the basis of description.

*Generalization in optical-Polarization.*-A plane monochromatic optical field propagating rectilinearly along z-direction in free space can, in general, be described by vector potential, $\vec{A}$ in the form,

$$\vec{A} = \hat{e}_x A_{0x} \cos(\psi - \phi_x) + \hat{e}_y A_{0y} \cos(\psi - \phi_y), \psi = \omega t - kz,$$

or in analytic-signal representation,

$$\vec{\mathcal{A}} = (\hat{e}_x \underline{A}_x + \hat{e}_y \underline{A}_y) e^{-i\psi}, \underline{A}_{x,y} = A_{0x,0y}\, e^{i\phi_{x,y}} \quad (1)$$

where $\underline{A}_{x,y}$ are classical complex amplitudes; $A_{0x,0y}$, real amplitudes and phase parameters, $\phi_{x,y}$ ($0 \leq \phi_{x,y} < 2\pi$) possess, in general, random spatio-temporal variation with angular frequency, $\omega$ in linear polarization basis $(\hat{e}_x, \hat{e}_y)$ of transverse plane, $\vec{k}$ (= $k\hat{e}_z$) is propagation vector of magnitude k, and $\hat{e}_{x,y,z}$ are unit vectors along respective x-, y-, z-axes forming right handed triad. Usually, optical-polarization is quantified by instantaneous Stokes theory. A simple classical argument may be placed to demonstrate inadequacy of Stokes theory at any spatial point. Stokes parameters are expressed as $s_{0,1}= \langle A_{0y}^2 \pm A_{0x}^2\rangle = \langle |\underline{A}_y|^2 \pm |\underline{A}_x|^2\rangle$; $s_2 + is_3 = 2\langle A_{0y}A_{0x}e^{i(\phi_y-\phi_x)}\rangle = 2\langle \underline{A}_y \underline{A}_x^*\rangle$, where the angular bracket, $\langle \rangle$ denotes the average over ensemble of realizations of complex amplitudes, $\underline{A}_{x,y}$ at any spatio-temporal points. Notably, unpolarized light is characterized by the values $s_0 \neq 0$, $s_1 = s_2 = s_3 = 0$. Let us consider an experiment in which amplitude-coherent, phase-randomized bi-modal optical beam propagating along z-axis (Eq. 1) is allowed to incident on a polarizer which extracts the linearly polarized component inclined at an angle $\theta$ with the x-axis, $\underline{A}_\theta = \underline{A}_x\cos\theta + \underline{A}_y\sin\theta$, followed by a non-linear crystal producing in its output second harmonic signal with a detector for the same. The detector would detect "square of intensity", proportional to $\langle \underline{A}_\theta^{*2} \underline{A}_\theta^2 \rangle$. Simple straightforward calculation reveals that $\langle \underline{A}_\theta^{*2} \underline{A}_\theta^2 \rangle = \frac{A_0^4}{4}[5-\cos 4\theta]$, where $A_{0y} = A_{0x} = A_0$ are taken, ensuring asymmetrical statistical properties (polarization structure) of considered optical field about direction of propagation, z-axis and, therefore, breaking cylindrical symmetry, for light to be unpolarized. But, if Stokes theory were applied to characterize the state of polarization, it would result in the unpolarized state due to random nature of phases, $\phi_{x,y}$ making all parameters to have vanishing values except $s_0$, a paradoxical tailspin. Evidently, Stokes theory misleads in assessing the state of optical-polarization, especially, when the higher-order correlations of optical-field variables are critical [37].

We contrived a primitive property of perfect polarized optical field, namely, the non-random ratio of complex amplitudes of transverse orthogonally polarized modes in, say, linear- polarization basis and defined it as 'index of polarization' (IOP) [34], i.e., $\underline{A}_y/\underline{A}_x = p$, a non-random complex parameter, which renders characteristic polarization - parameters ('ratio of real amplitudes' and 'difference in phases'). Clearly, polarized optical field (i.e., non random, p is known) is a mono - modal since only one random complex amplitude suffices for its complete statistical description (other orthogonal complex amplitude is specified through p). Moreover, if one introduces new parameters, $A_0$ (real random amplitude defining global intensity), $\chi_0$ (polar angle), $\Delta_0$ (azimuth angle), $\phi$ (random global phase) on a Poincare sphere, satisfying inequalities $0 \leq A_0$, $0 \leq \chi_0 \leq \pi$, $-\pi < \Delta_0 \leq \pi$, $0 \leq \phi < 2\pi$, respectively, involving transforming equations in terms of old parameters, $A_0 = (A_{0x}^2 + A_{0y}^2)^{1/2}$, $\chi_0 = 2\tan^{-1}(A_{0y}/A_{0x})$ and $\Delta_0 = \phi_y - \phi_x$; $\phi = (\phi_x + \phi_y)/2$, the analytical signal, $\vec{\mathcal{A}}$, in Eq. (1), yields a self-instructive form,

$$\vec{\mathcal{A}} = \hat{\varepsilon}_0 \mathcal{A}; \quad \mathcal{A} = \underline{A}e^{-i\psi}; \quad \underline{A} = A_0 e^{i\phi},$$
$$\hat{\varepsilon}_0 = \hat{e}_x \cos\frac{\chi_0}{2} e^{-\Delta_0/2} + \hat{e}_y \sin\frac{\chi_0}{2} e^{\Delta_0/2}. \tag{2}$$

The form of vector potential, $\vec{\mathcal{A}}$ in Eq. (2) may be interpreted as a single mode polarized optical field, statistically explicable by single complex amplitude, $\underline{A}$ polarized in the fixed direction, $\hat{\varepsilon}_0$ determined by non-random angle parameters $\chi_0$ and $\Delta_0$ on the Poincare sphere specifying the polarization mode, $(\hat{\varepsilon}_0, \vec{k})$ and demarcating itself from spatio-temporal mode $\psi = \omega t - kz$. Here, the complex vector $\hat{\varepsilon}_0$ is a unit vector ($\hat{\varepsilon}_0^* \cdot \hat{\varepsilon}_0 = 1$) giving expression of IOP on Poincare sphere, $p = \underline{A}_y/\underline{A}_x = \tan\frac{\chi_0}{2} e^{i\Delta_0}$. Visibly, the state of optical-polarization is specified by the non-random values of p, which, in turn, is fixed by non-random values of $\chi_0$ and $\Delta_0$ defining a point ($\hat{\varepsilon}$), on the unit Poincare sphere, similar to its counterpart Stokes parameters. All typical ellipsometrical-parameters such as orientation angle and eccentricity of the polarization-ellipse can be determined if p of optical field and one complex amplitude are specified.

The linear polarization basis, ($\hat{e}_x$, $\hat{e}_y$) doesn't enjoy privileged position in conventional theory as the elliptic-polarization basis ($\hat{\varepsilon}$, $\hat{\varepsilon}_\perp$), under suitable SU(2) polarization transformations [38], may be degenerated into linear-polarization basis ($\hat{e}_x$, $\hat{e}_y$) and circular polarization basis ($\hat{e}_R$, $\hat{e}_L$), $\hat{e}_{R,L} =$ $\frac{1}{\sqrt{2}}(\hat{e}_x \pm i\,\hat{e}_y)$. Also, the elliptic-polarization basis, $(\hat{\varepsilon}, \hat{\varepsilon}_\perp)$ facilitates generality in the formulation in which polarized light (Eq. 2) retain IOP, $p_{(\hat{\varepsilon},\hat{\varepsilon}_\perp)}$, another non-random parameter showing vivid dependence on $\hat{\varepsilon}_0$,

$$p_{(\hat{\varepsilon},\hat{\varepsilon}_\perp)} = \underline{A}_{\hat{\varepsilon}_\perp}/\underline{A}_{\hat{\varepsilon}} = \frac{\hat{\varepsilon}_\perp^*\cdot\hat{\varepsilon}_0}{\hat{\varepsilon}^*\cdot\hat{\varepsilon}_0}, \qquad (3)$$

where $\hat{\varepsilon}_\perp$ is orthogonal complex unit vector ($\hat{\varepsilon}_\perp^*\cdot\hat{\varepsilon}_\perp = 1$, $\hat{\varepsilon}_\perp^*\cdot\hat{\varepsilon} = 0$). This formula caters interchangeability of IOP's between different polarization bases of descriptions. Polar parameterization of complex amplitudes $\underline{A}_{\hat{\varepsilon}}$ and $\underline{A}_{\hat{\varepsilon}_\perp}$ by $\underline{A}_{\hat{\varepsilon}} = A_{0\hat{\varepsilon}} \exp(i\phi_{\hat{\varepsilon}})$, $\underline{A}_{\hat{\varepsilon}_\perp} = A_{0\hat{\varepsilon}_\perp} \exp(i\phi_{\hat{\varepsilon}_\perp})$, with real-amplitudes $(A_{0\hat{\varepsilon}}, A_{0\hat{\varepsilon}_\perp})$ and phase-parameters $(\phi_{\hat{\varepsilon}}, \phi_{\hat{\varepsilon}_\perp})$, the Eq. (3) reveals that a polarized light maintain non-random values of (i) 'ratio of real amplitudes', $A_{0\hat{\varepsilon}_\perp}/A_{0\hat{\varepsilon}}$, and (ii) 'difference in phases', $(\phi_{\hat{\varepsilon}_\perp} - \phi_{\hat{\varepsilon}})$, deciphering characteristic polarization parameters. Recently, one of the authors and others [39] studied the generation of Hidden optical-polarization, by spatially structured bi-modal coherent light as a pump, in degenerate parametric down conversion in which instead of non-random values of characteristic polarization parameters one has non-random values of (i) 'ratio of real amplitudes' and 'sum of phases' contrary to 'difference in phases' for usual polarized light. The Stokes theory for such kind of optical field assigns the status of the unpolarized state which is shown not to be true fact because it is a mono-modal optical field with peculiar transversal structure.

The concept of HOOP is introduced by demanding non-random values of all multiple-exponents of some positive integer, say, n of the ratio of complex amplitudes unless it takes unit value. The IOP, characterizing light in $n^{th}$-order optical-polarization, may be termed as,

$$p_{(\hat{\varepsilon},\hat{\varepsilon}_\perp),n} \equiv (\underline{A}_{\hat{\varepsilon}_\perp}/\underline{A}_{\hat{\varepsilon}})^n. \qquad (4)$$

The positive integer, n takes unit value for usual polarized light (cf. eq. 3). For HOOP, where n is necessarily greater than unity, the random value of the ratio, $\underline{A}_{\hat{\varepsilon}_\perp}/\underline{A}_{\hat{\varepsilon}}$ and non-random value of $n^{th}$-exponent of the ratio, $(\underline{A}_{\hat{\varepsilon}_\perp}/\underline{A}_{\hat{\varepsilon}})^n = p_{(\hat{\varepsilon},\hat{\varepsilon}_\perp),n} = |p_{(\hat{\varepsilon},\hat{\varepsilon}_\perp),n}| \exp(i\Delta_{(\hat{\varepsilon},\hat{\varepsilon}_\perp),n})$, demonstrates that while ratio of real amplitudes, $A_{0\hat{\varepsilon}_\perp}/A_{0\hat{\varepsilon}}$ has a non-random value, $|p_{(\hat{\varepsilon},\hat{\varepsilon}_\perp),n}|^{1/n}$, difference in phases, $\phi_{\hat{\varepsilon}_\perp} - \phi_{\hat{\varepsilon}}$ may have equally likely values among 'n' non-random values, $\frac{1}{n}(\Delta_{(\hat{\varepsilon},\hat{\varepsilon}_\perp),n} + 2r\pi)$ with r = 0, 1, 2,…(n-1), in steps of $(2\pi/n)$. Evidently, these many values of phases make HOOP to abide by weird polarization nature and, hence, not a mono-modal optical field contrary to the conventional perfect polarized light which is ascertained by single non-random ratio of real amplitudes and non-random difference in phases in the elliptic-polarization basis $(\hat{\varepsilon}, \hat{\varepsilon}_\perp)$.

One may develop quantum theory of HOOP on a similar classical lineage. In Quantum Optics the optical field, Eq. (1) is described by operatic-version of vector potential operator, $\vec{\mathcal{A}} = \left(\frac{2\pi}{\omega V}\right)^{1/2} [(\hat{e}_x \hat{a}_x + \hat{e}_y \hat{a}_y)e^{-i\psi} + \text{h.c.}] = \left(\frac{2\pi}{\omega V}\right)^{1/2} [(\hat{\varepsilon}\hat{a}_{\hat{\varepsilon}} + \hat{\varepsilon}_\perp \hat{a}_{\hat{\varepsilon}_\perp})e^{-i\psi} + \text{h.c.}]$, in linear-polarization basis $(\hat{e}_x, \hat{e}_y)$ or in elliptic-polarization basis $(\hat{\varepsilon}, \hat{\varepsilon}_\perp)$, respectively, where $\omega$ is angular frequency of the optical field and V is the quantization volume of the cavity, h.c. stands for Hermitian conjugate. Using orthonormal properties of complex unit vectors $\hat{\varepsilon} (= \varepsilon_x \hat{e}_x + \varepsilon_y \hat{e}_y)$ and $\hat{\varepsilon}_\perp (= \varepsilon_{\perp x}\hat{e}_x + \varepsilon_{\perp y}\hat{e}_y)$ the annihilation operators $\hat{a}_{\hat{\varepsilon}}$ ($\hat{a}_{\hat{\varepsilon}_\perp}$) are related with those in linear-polarization basis $(\hat{e}_x, \hat{e}_y)$ by the expressions $\hat{a}_{\hat{\varepsilon}} = \varepsilon_x^* \hat{a}_x + \varepsilon_y^* \hat{a}_y$, $\hat{a}_{\hat{\varepsilon}_\perp} = \varepsilon_{\perp x}^* \hat{a}_x + \varepsilon_{\perp y}^* \hat{a}_y$, satisfying usual Bosonic-commutation relations.

The pure dynamical state of a monochromatic optical beam, propagating along z-axis and polarized in the mode, $(\hat{\varepsilon}_0, \vec{k})$, may be specified by a state vector $|\psi\rangle$ in Hilbert space. Evidently, such light doesn't have signal (photons) in orthogonal mode $(\hat{\varepsilon}_{0\perp}, \vec{k})$, i.e.,

$$\hat{a}_{\hat{\varepsilon}_{0\perp}} |\psi\rangle = 0, \tag{5}$$

which yields, on applying the above expression for $\hat{a}_{\hat{\epsilon}_{0\perp}}$, $(\epsilon^*_{0\perp x} \hat{a}_x + \epsilon^*_{0\perp y} \hat{a}_y)|\psi\rangle = 0$. Refurbishing by orthogonality relation between $\hat{\epsilon}_0$ and $\hat{\epsilon}_{0\perp}$, one obtains the defining equation for perfect optical-polarization,

$$\hat{a}_y |\psi\rangle = p\hat{a}_x|\psi\rangle, \qquad (6)$$

which is the quantum analogue to the classical perfect polarization criterion, $\underline{A}_y = p\underline{A}_x$, giving p = $\epsilon_{0y}/\epsilon_{0x}$ (= $\tan\frac{\chi_0}{2} e^{i\Delta_0}$, from Eq. 2). Casting Eq. (6) in elliptic-polarization basis $(\hat{\epsilon}, \hat{\epsilon}_\perp)$, yields $\hat{a}_{\hat{\epsilon}_\perp}|\psi\rangle = p_{(\hat{\epsilon},\hat{\epsilon}_\perp)} a_{\hat{\epsilon}}|\psi\rangle$, upon which repetitive applications by annihilation operator 'r'-times (r, any positive integer) reproduces,

$$(\hat{a}_{\hat{\epsilon}_\perp})^r|\psi\rangle = p_{(\hat{\epsilon},\hat{\epsilon}_\perp),r} (a_{\hat{\epsilon}})^r|\psi\rangle. \qquad (7)$$

We set up the quantum version of the classical criterion for HOOP (see Eq. 4) by demanding,

$$(\hat{a}_{\hat{\epsilon}_\perp})^n|\psi\rangle = p_{(\hat{\epsilon},\hat{\epsilon}_\perp),n}(a_{\hat{\epsilon}})^n|\psi\rangle, \qquad (8)$$

for the least multiple value n of the positive integer, r. It is evident that for n = 1, usual optical-polarization results (cf. Eq. 6) revealing that conventional theory of optical-polarization is first-order HOOP. For n > 1, defining criterion for HOOP, Eq. (8) is fulfilled only when n is the least multiple of positive integer r. On the other hand for mixed state of optical field described by the density operator, $\hat{\rho}$, prescription, Eq. (8) assumes the form, $(\hat{a}_{\hat{\epsilon}_\perp})^n \hat{\rho} = p_{(\hat{\epsilon},\hat{\epsilon}_\perp),n}(\hat{a}_{\hat{\epsilon}})^n \hat{\rho}$. Expressing density operator $\hat{\rho}$ in Glauber-Sudarshan P-representation [40], $\hat{\rho} = \int d^2\alpha\, d^2\beta\, P(\alpha,\beta)|\alpha,\beta\rangle_{(\hat{\epsilon},\hat{\epsilon}_\perp)}\, _{(\hat{\epsilon},\hat{\epsilon}_\perp)}\langle\alpha,\beta|$, in the basis of bi-modal quadrature coherent states, $|\alpha,\beta\rangle_{(\hat{\epsilon},\hat{\epsilon}_\perp)}$ satisfying familiar eigenvalue equations, $(\hat{a}_{\hat{\epsilon}}, \hat{a}_{\hat{\epsilon}_\perp})|\alpha,\beta\rangle_{(\hat{\epsilon},\hat{\epsilon}_\perp)} = (\alpha,\beta)|\alpha,\beta\rangle_{(\hat{\epsilon},\hat{\epsilon}_\perp)}$, where $\alpha$, $\beta$ are classical complex amplitudes along the polarization modes $(\hat{\epsilon}, \vec{k})$ and $(\hat{\epsilon}_\perp, \vec{k})$ respectively, we obtain quasi-probability function, $P(\alpha,\beta) = \delta^2(\beta^n - p_{(\hat{\epsilon},\hat{\epsilon}_\perp),n}\alpha^n)$, two-dimensional Dirac-delta function, defined for a complex number z = x + iy by expression, $\delta^2(z) = \delta^2(x)\delta^2(y)$, which provides IOP, $p_{(\hat{\varepsilon},\hat{\varepsilon}_\perp),n} = \beta^n/\alpha^n$. Writing IOP, p in polar form, $p_{(\hat{\varepsilon},\hat{\varepsilon}_\perp),n} = |p_{(\hat{\varepsilon},\hat{\varepsilon}_\perp),n}|\exp(i\Delta_{(\hat{\varepsilon},\hat{\varepsilon}_\perp),n})$ and polar decompositions of classical complex amplitudes, α, β, one may extract the same property of HOOP as in classical theory, i. e., the ratio of real amplitudes has single non-random value, $|p_{(\hat{\varepsilon},\hat{\varepsilon}_\perp),n}|^{1/n}$ and the phase difference might pick up any one of the 'n' non-random values $\frac{1}{n}(\Delta_{(\hat{\varepsilon},\hat{\varepsilon}_\perp),n} + 2r\pi)$ with r = 0, 1,… (n-1) in elliptic polarization basis, $(\hat{\varepsilon},\hat{\varepsilon}_\perp)$.

Typical instances of HOOP-states.- Stokes seems to be incapable in quantifying polarization properties of bi-modal optical non-Guassian Schrodinger Cat states or Cat-like states and experimentally generated entangled bi-modal coherent states [41]. If bimodal optical field is prepared as to be in coherent state in one mode and other mode is in odd or even coherent state [42], i.e., if $|\psi\rangle_\pm \propto |\alpha\rangle_x(|\beta\rangle \pm |-\beta\rangle)_y$, one obtains, $\hat{a}_y^2|\psi\rangle_\pm = (\beta^2/\alpha^2)\hat{a}_x^2|\psi\rangle_\pm$, similar to Eq. (8) but not relation of the form, $\hat{a}_y|\psi\rangle_\pm = p\hat{a}_x|\psi\rangle_\pm$ for usual perfect polarized light (see Eq.6). So, the optical field, $|\psi\rangle_\pm$ is attributed to possess second-order HOOP in the linear polarization basis $(\hat{e}_x,\hat{e}_y)$ with IOP, $p_{(\hat{e}_x,\hat{e}_y),2}$ ($\equiv \beta^2/\alpha^2$). Examples of still higher HOOP states of light are provided when the optical fields are in $|\psi\rangle_\pm \propto |\alpha\rangle_x(|\beta\rangle \pm |\beta^*\rangle)_y$, *if $\beta^n$ is real*, one gets $\hat{a}_y^n|\psi\rangle_\pm = (\beta^n/\alpha^n)\hat{a}_x^n|\psi\rangle_\pm$ and, therefore, $|\psi\rangle_\pm$ are ascertained to be polarized in the n$^{th}$ order in the basis $(\hat{e}_x,\hat{e}_y)$ with IOP, $p_{(\hat{e}_x,\hat{e}_y),n}$ ($\equiv \beta^n/\alpha^n$). Moreover, if both polarization modes, $\hat{\varepsilon}$ and $\hat{\varepsilon}_\perp$ components are Cat states or Cat-like states, $|\psi\rangle = \left(\sum_{s_1=0}^{n_1-1} c_{s_1}|\alpha e^{i2\pi s_1/n_1}\rangle\right)_{\hat{\varepsilon}} \left(\sum_{s_2=0}^{n_2-1} d_{s_2}|\beta e^{i2\pi s_2/n_2}\rangle\right)_{\hat{\varepsilon}_\perp}$, it displays n$^{th}$ order polarization, where n being least common multiple of both positive integers $n_1$ and $n_2$.

Writing entangled bi-modal coherent states [41] in a linear polarization basis $(\hat{e}_x,\hat{e}_y)$, $|\psi_\phi\rangle = -i\sin\frac{\phi}{2}|\alpha,\alpha\rangle_{(x,y)} - \cos\frac{\phi}{2}|-\alpha,\alpha\rangle_{(x,y)} + \cos\frac{\phi}{2}|\alpha,-\alpha\rangle_{(x,y)} + i\sin\frac{\phi}{2}|-\alpha,-\alpha\rangle_{(x,y)}$, where $|\alpha,\alpha\rangle_{(x,y)} = |\alpha\rangle_x|\alpha\rangle_y$, etc., we obtain $\hat{a}_y^2|\psi_\phi\rangle = \hat{a}_x^2|\psi_\phi\rangle$, revealing that $|\psi_\phi\rangle$ is in 2$^{nd}$- order HOOP.

The 'statement of the basis' in theory of HOOP offers a pivotal point which, in contrast, imparts a trivial implication in theory of usual optical-polarization [43]. For HOOP light in one basis may not be even polarized in any order in some other basis, or may be polarized with different order in some variant basis. As a revelation of this nature, one may observe that light, being in the pure state $|\psi\rangle \propto |i\sqrt{3}\,\alpha\rangle_x (|\alpha\rangle + |-\alpha\rangle)_y$ and satisfying $\hat{a}_y^2|\psi\rangle = (-1/3)\,\hat{a}_x^2|\psi\rangle$, is polarized in the 2$^{nd}$-order with IOP, $p = -1/3$ in the basis $(\hat{e}_x, \hat{e}_y)$, while in the polarization basis $\hat{e}_\pm = (\hat{e}_y \pm \hat{e}_x)/\sqrt{2}$, following Heisenberg convention and the expression, $\hat{a}_-^3 - \hat{a}_+^3 = 3a_x(\hat{a}_y^2 - \frac{1}{3}\hat{a}_x^2)$, one obtains $\hat{a}_-^3|\psi\rangle = \hat{a}_+^3|\psi\rangle$ which displays polarization in 3$^{rd}$-order with polarization index 1 in the basis $(\hat{e}_+, \hat{e}_-)$. The genesis of huge dependence of order (n$^{th}$) on bases of polarization, i.e., existence of preferential basis for its manifestation seems to be a peculiar feature of the HOOP which, intimately, originates from many (n) values of non-random phases revealing that HOOP light is an example of pseudo mono-modal or multi-modal optical field.

In conclusion, the concept of Higher-order polarization of light is introduced by generalizing the basic property of perfect optical-polarization, i.e., the least multiple-exponent, say, positive integer n raised as to the power of ratio of complex amplitudes of polarized components in transverse orthogonal modes to be non-random parameter. The efficacy of criterion for HOOP (see Eq.8) is exemplary as it deciphers not only characteristic polarization parameters of light but also discerns the higher-order polarization nature. The criterion of HOOP presents a simple computable characterization measure applicable to almost each bimodal quantum states of light. The transition of quantum states revealing HOOP from Hilbert space to quantum phase space highlights the critical importance of its pseudo mono-modal or multi-modal character, which will be taken as a resource for Higher-order Wigner distribution function foisted on polarization description of entangled bi-modal coherent states [44].

We acknowledge the illuminating discussions with Professors N. Chandra and R. Prakash, University of Allahabad, Allahabad, India. One of the authors (RSS) gratefully indebted to Prof.

[#]yesora27@gmail.com

along z-direction. We obtain the same statistical property (square of the intensity) rendering polarization structure as Tr[$\hat{\rho}\,\hat{a}_\theta^{\dagger 2}\,\hat{a}_\theta^{2}$] = $\frac{A_0^4}{4}$ [5-cos 4θ], provided that $\hat{a}_\theta \equiv \hat{a}_x \cos\theta + \hat{a}_y \sin\theta$. But, on account of random nature of phases of $\alpha_{x,y}$, Stokes parameters $s_1, s_2, s_3$ are all zero except $s_0$.